\documentclass[12pt]{article}
\usepackage[a4paper]{geometry}
\usepackage{graphicx}

\usepackage{amsthm}
\usepackage{amsmath}
\usepackage{amssymb}
\usepackage{amsfonts}
\usepackage{graphicx}
\usepackage{graphics}
\usepackage[bf,SL,BF]{subfigure}

\def\eref#1{(\ref{#1})}

\def\<{\big\langle}
\def\>{\big\rangle}

\theoremstyle{remark}

\theoremstyle{definition}

\newcommand{\figbox}[1]{%
  \fbox{%
    \vbox to 1in{%
    \vfil
    \hbox to 2in{%
      \hfil
      #1%
      \hfil}%
    \vfil}}}

\newcommand{\goodgap}{%
  \hspace{\subfigcapskip}}

\begin{document}
\title{Bistable equilibrium points of mercury body burden.}
\author{Houman Owhadi\footnote{California Institute of Technology Applied \&
Computational Mathematics, Control \& Dynamical systems, MC 217-50
Pasadena , CA 91125, owhadi@caltech.edu} and Areen
Boulos\footnote{areen@caltech.edu}}
 \maketitle

\begin{abstract}
In the last century mercury levels in the global environment have
tripled as a result of increased pollution from industrial,
occupational, medicinal and domestic uses \cite{BaMe03}. Glutathione
is known to be the main agent responsible for the excretion of
mercury (we refer to \cite{Thim05}, \cite{ZalBar99} and
\cite{Lyn02}). It has also been shown that mercury inhibits
glutathione synthetase (an enzyme acting in the synthesization of
Glutathione), therefore leading to decreased glutathione levels
 (we refer to \cite{Thim05}, \cite{GeGe05}, \cite{GeGe06} and \cite{RDeth04}).
Mercury also interferes with the production of heme in the porphyrin
pathway \cite{WoMaEc93}. Heme is needed for biological energy
production and ability to detox organic toxins via the P450 enzymes
\cite{Boy06}. The purpose of this paper is to show that body's
response to mercury exposure is hysteretic, i.e.  when this feedback
of mercury on its main detoxifying agents is strong enough then
mercury body burden has two points of equilibrium: one with normal
abilities to detoxify and low levels of mercury and one with
inhibited abilities to detoxify and high levels of mercury.
Furthermore, a small increase of body's mercury burden may not be
sufficient to trigger observable neurotoxic effects but it may be
sufficient to act as a switch leading to an accumulation of mercury
in the body through environmental exposure until its toxicity
becomes manifest.
\end{abstract}

\section{The model}

Writing $C$ the mercury body burden a simple model for the evolution
of $C$ with respect to time can be described by the following
ordinary differential equation.
\begin{equation}\label{eq1}
\frac{dC}{dt}=\mu-k(C) C
\end{equation}
In equation \eref{eq1}, $\mu$ stands for the rate of exposure of the
body to mercury.  $k(C) C$ stands for the detoxification rate. As an
instance of chronic exposure, recall that the intra oral air
concentration in mercury is of the order of $2 \mu g/m^3$ without
dental amalgam and $20 \mu g/m^3$ with dental amalgams \cite{VL85}.
It has also been shown that dental amalgam filings in pregnant rats
can be a prime source of exposure of fetal tissues to mercury
\cite{TaTsHa01}.

Although equation \eref{eq1} is a toy model we believe that it
captures the main mechanism involved in mercury exposure. This toy
model is based on the following assumptions:
\begin{itemize}
\item The excretion rate of mercury is proportional to the mercury
body burden and we write $k(C)$ the proportionality constant.
\item The proportionality constant $k(C)$ is a decreasing function
of the mercury burden $C$ modeling the inhibition of organs ability
to detoxify by mercury.
\end{itemize}

The model introduced in this paper is oversimplified but we believe
that the associated switch mechanism  could be of some relevance in
the analysis of mercury toxicity.

Observe that one has to distinguish two main factors in mercury
toxicity. The first one (addressed in this paper) is the total
amount of mercury accumulated in the body. The second one (not
addressed in this paper), is the mercury retention effects, i.e. the
toxicity of a given value of mercury body burden (see \cite{Boy06b},
\cite{NeurotoxEch05}, \cite{NeurotoxEch06} and the references
therein).

\begin{itemize}
\item Low levels of mercury lead to aberrant porphyrin
profile, the major product of the porphyrin pathway is heme and it
is needed for biological energy production and ability to detoxify
organic toxins via the P450 enzymes.

\item Genetic susceptibilities that account for susceptibility
  to heavy metal toxicities (such as polymorphism in the CPOX4 or APO-E genes) are an important
  factor affecting excretion rates.

\item Genetic variations of Apolipoprotein E (APO-E) (\cite{Boy06}, \cite{CoSa93},
\cite{GodWo03}) don't play  a major role in mercury excretion but
play a role in mercury retention effects by altering the
detoxification abilities of the Central Nervous System by impacting
the percentage of the total mercury body burden trapped in the
cerebrospinal system. Hence although being an APO-E4  carrier won't
significantly impact excretion rates it increases the susceptibility
Alzheimer's disease  (\cite{Boy06}, \cite{GodWo03}) by increasing
the relative percentage of mercury withheld in the central nervous
system.
\end{itemize}

This being said, it is important to observe that although genetic
variations may alter excretion rates ($k(C)$ would in practice
depend on genetic susceptibilities), it would not explain a
feed-back mechanism. Mercury is primarily excreted bound to
glutathione (\cite{Thim05}, \cite{ZalBar99} and \cite{Lyn02}) and
mercury can inhibit the production of glutathione \cite{RDeth04}
(\cite{Thim05}, \cite{GeGe05}, \cite{GeGe06} and \cite{RDeth04}). We
will show that through this feedback mechanism,  very small doses of
mercury cause a major breakdown of the detoxification pathway.

\begin{figure}[httb]
  \begin{center}
    \subfigure[$k(C)$ versus $C$.]
    {\includegraphics[width=0.45\textwidth,height= 0.4\textwidth]{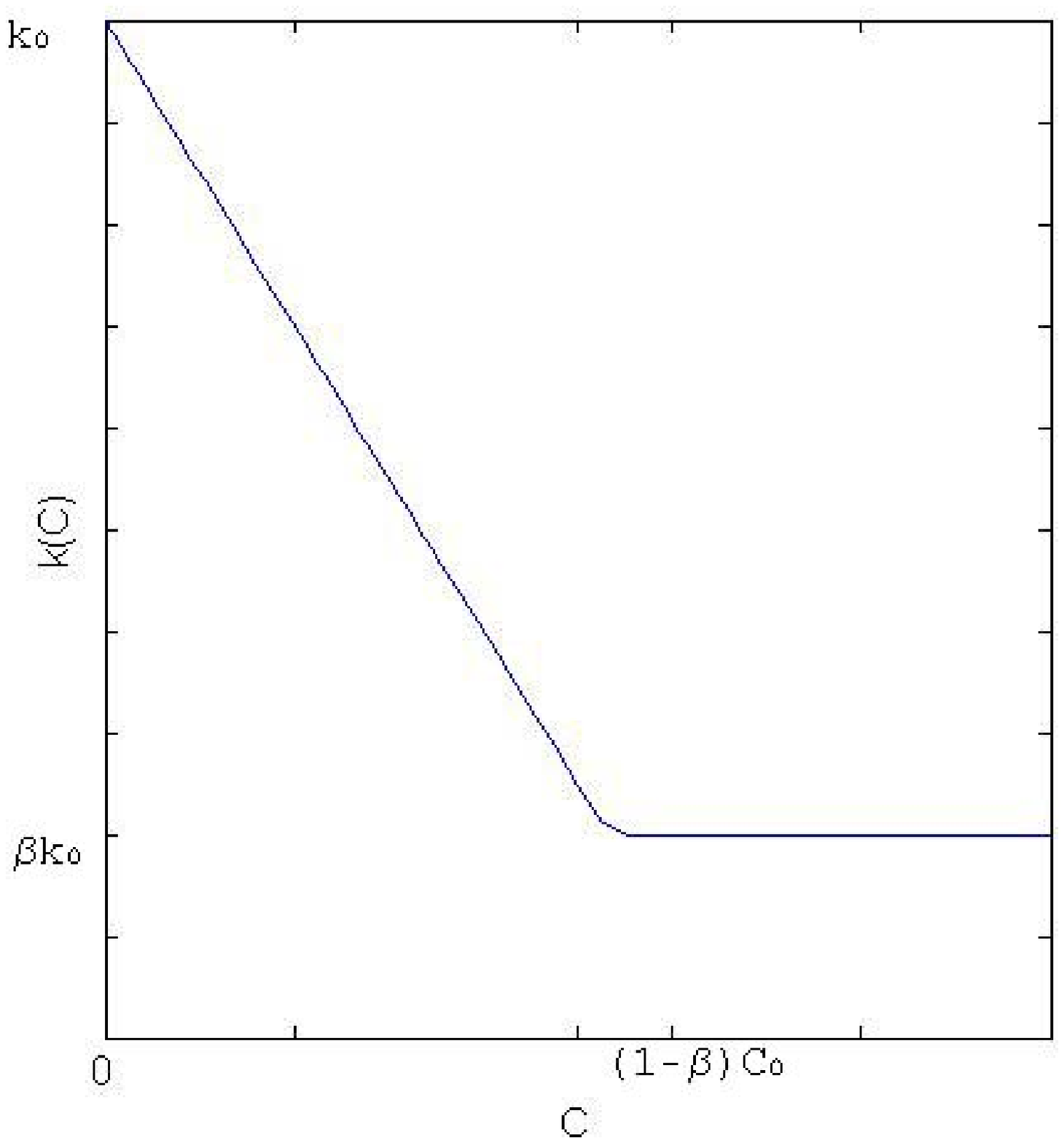}\label{figdeerr1}}
    \goodgap
    \subfigure[$k(C)C$ versus $C$.]
    {\includegraphics[width=0.45\textwidth,height= 0.4\textwidth]{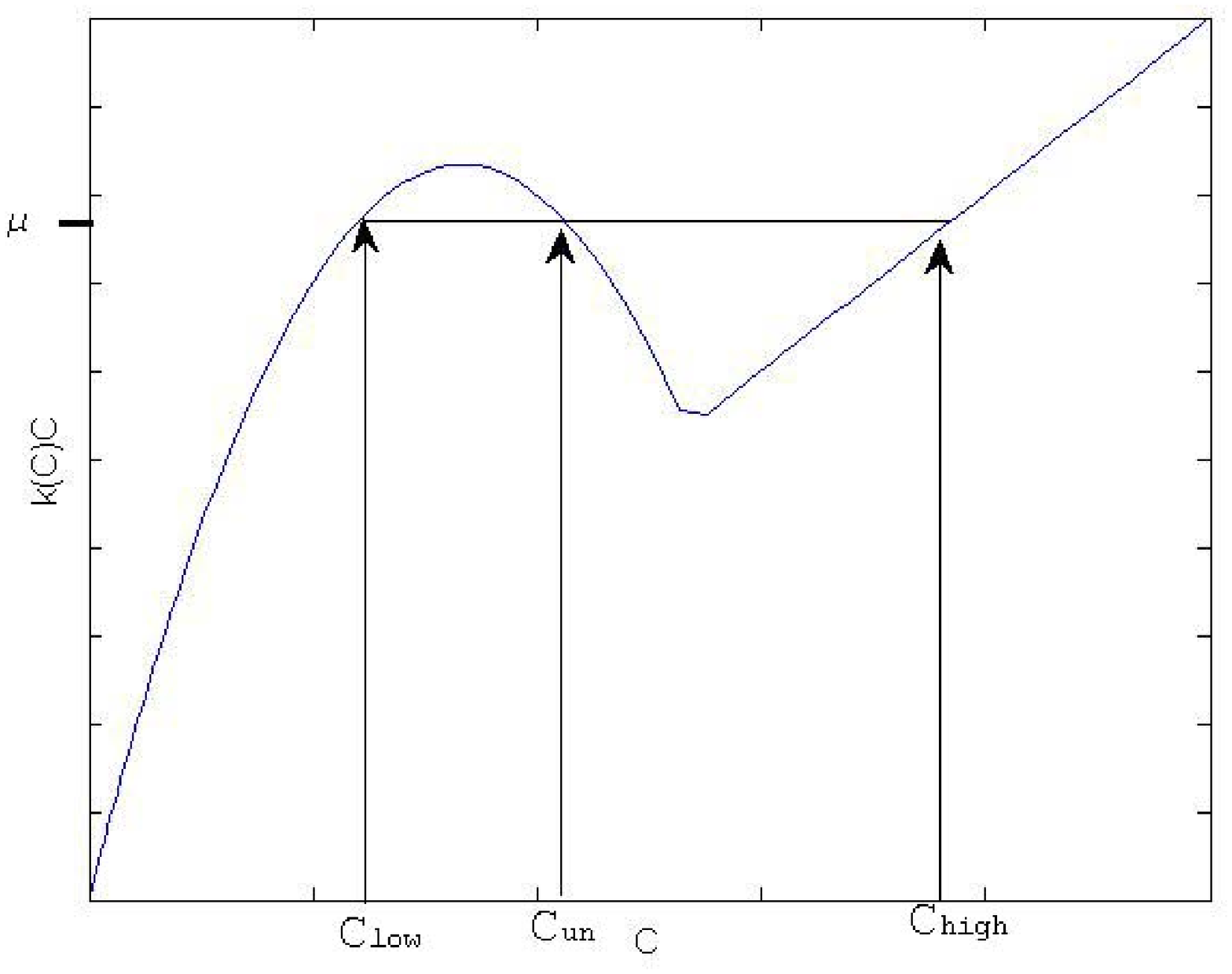}\label{kccccddc}}\\
    \caption{$k(C)$ and $k(C)C$ versus $C$.}
    \label{unut100te44ip7}
\end{center}
\end{figure}

To simplify the presentation we will assume that $k(C)$ has the
following form
\begin{equation}\label{eq2}
k(C)= k_0 \max(1-\frac{C}{C_0},\beta)
\end{equation}
where $k_0$ is the detoxification rate at low levels of mercury and
$\beta$ is a constant between $0$ and $1$ indicating the decrease of
the proportionality constant $k(C)$ under levels of mercury beyond
$C_0(1-\beta)$. Thus $C_0$ is a positive parameter and
$C_0(1-\beta)$ is a constant indicating a critical level of mercury
beyond which Glutathione production is minimal. Figure
\ref{figdeerr1} shows how $k(C)$ would vary as a function of $C$ in
this model.

 One
should observe that the phenomenon described here is independent of
the particular choice of $k(C)$ and is solely based on the non
monotonicity of $k(C) C$ with respect to $C$ as in figure
\ref{kccccddc}.

The equilibrium points for body mercury burden are solutions of the
following equation
\begin{equation}\label{eq3}
k(C) C=\mu
\end{equation}
More precisely
\begin{equation}\label{eq4}
\max(1-\frac{C}{C_0},\beta) \frac{C}{C_0}=\frac{\mu}{k_0 C_0}
\end{equation}
Equation \eref{eq4} has three solutions if an only if
\begin{equation}\label{eq5}
\beta (1-\beta)< \frac{\mu}{k_0 C_0}<\frac{1}{4}
\end{equation}
Let us write $C_{low}< C_{unstable}< C_{high}$ those equilibrium
points. Thus the dynamic of the body mercury burden can be divided
into two classes.

\begin{figure}[httb]
  \begin{center}
    \subfigure[Direction fields of mercury body burden.]
    {\includegraphics[width=0.45\textwidth,height= 0.4\textwidth]{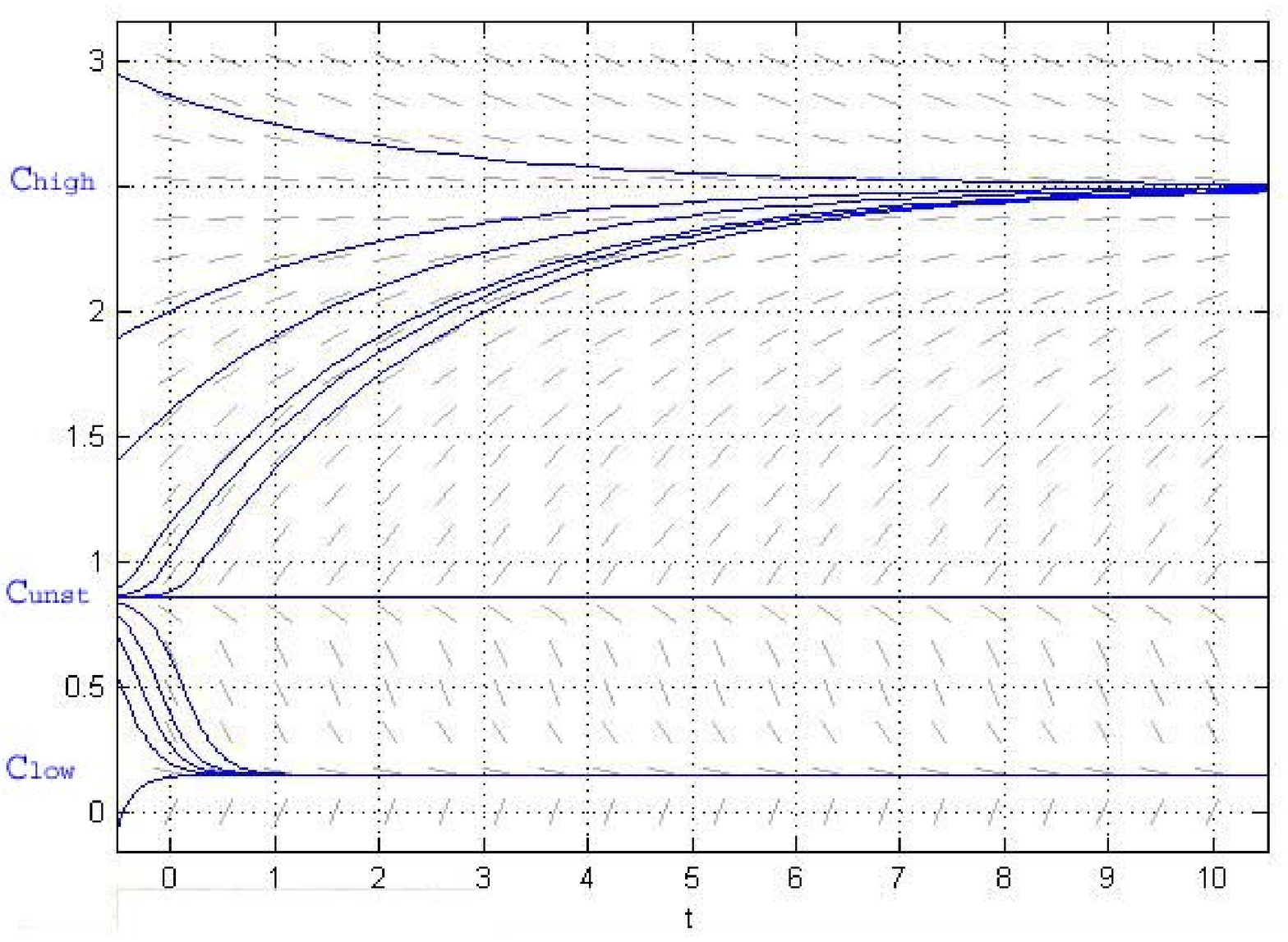}\label{fig1}}
    \goodgap
    \subfigure[Body burden is initially $0$ with a low rate of exposure then at time $2$ the body burden is
    increased in one moment.]
    {\includegraphics[width=0.45\textwidth,height= 0.4\textwidth]{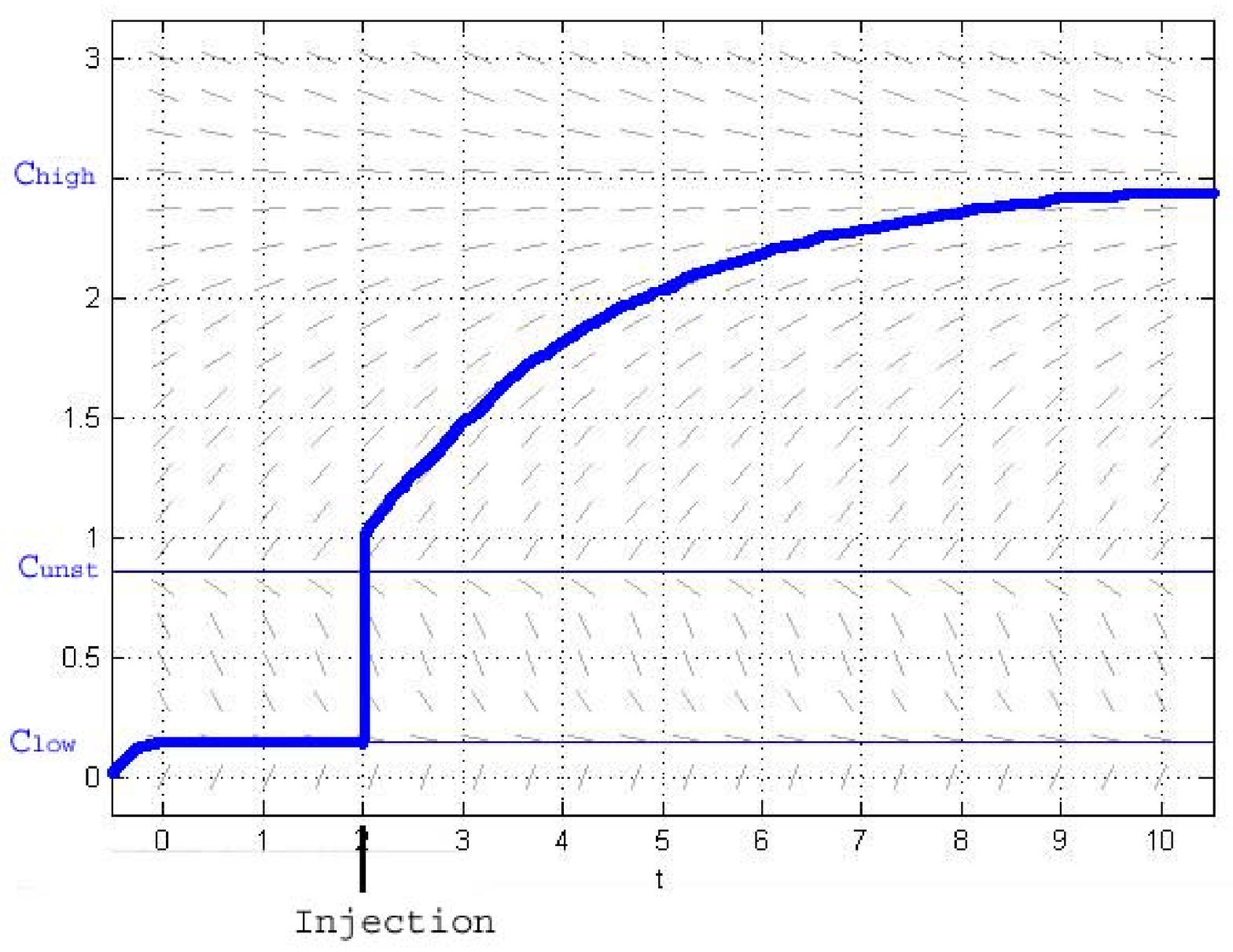}\label{fig2}}\\
    \caption{Mercury burden dynamic.}
    \label{unut100tip7}
\end{center}
\end{figure}

\paragraph{Unique equilibrium point.} This class is characterized by
the fact that the chronic exposure to mercury $\mu$ is above the
value $\frac{k_0 C_0}{4}$ or below the value $\beta (1-\beta) k_0
C_0$. In this dynamic $C(t)$ will converge to the unique solution of
\eref{eq4} as $t\rightarrow \infty$.

\paragraph{Bistable equilibrium points.}
This class is characterized by the fact that $\beta <\frac{1}{2}$
(the decrease of $k(C)$ under high levels of mercury is at least
$50\%$) and that  the chronic exposure to mercury $\mu$ is below the
value $\frac{k_0 C_0}{4}$ and above the value $\beta (1-\beta) k_0
C_0$. In this case if $C(0)\in (0, C_{unstable})$ then as
$t\rightarrow \infty$, $C(t)$ will converge to $C_{low}$. If
$C(0)>C_{unstable}$ then $C(t)$ will converge to $C_{high}$ as
$t\rightarrow \infty$. $C_{unstable}$ is an unstable equilibrium
point. We refer to figure \ref{fig1}. Observe also that in this
situation a high mercury body burden can be induced by exposing the
organism to a given small amount of mercury in a short lapse of time
(see figure \ref{fig2}).

\paragraph{Experimental detoxification rates.}
It would be interesting to obtain constant rates that apply to
elemental Hg, inorganic Hg, and organic types of Hg. While
\cite{AtAbsHg83} contains excretion rate those rates needs to be put
into correspondence with total body burden to infer the experimental
estimation of the curve $k(C)C$. Such experimental data would be
very valuable since it would allow us to compute the value
$C_{unstable}-C_{low}$ (which may strain dependent) which is the
amount of mercury leading  to an activation of the switch mechanism
if the body is exposed to it in one time.

Observe also that one could elaborate more sophisticated models
where different organs have different detoxification rate (depending
as well upon the chemical form of the mercury). Such models would
put into evidence the biphasic blood detoxification rates for methyl
mercury (with average half periods of 7-8 hours and 52 days). It is
important to observe that the mechanism involved in the latter
biphasic behavior (trapping of methyl-mercury by organs which are
slow to detoxify) is different from one involved in this paper
(inhibition of the ability to detoxify).

\section{Interpretation and predictions.} According to the model
presented in this paper,  a transition from the low levels of
mercury body burden to the high levels of mercury body burden can be
induced by increasing the mercury body burden in a brief period of
time (increasing $C$ through in a brief period of time) or
increasing the rate of exposure ($\mu$) over a long enough period of
time. It is important to observe that to reach high levels of
mercury body burden $C_{high}$ it is enough to increase the body
burden by an amount $C_{unstable}-C_{low}$ which can be much smaller
than $C_{high}-C_{low}$ (see figure \ref{fig2}).

Observe that a prediction of our model is that once the metabolism
is in the basin of attraction of  inhibited abilities to detoxify it
will start accumulating not only mercury, but also other heavy
metals. Furthermore to induce a transition from inhibited abilities
to detoxify to the basin of attraction of normal  abilities to
detoxify one would have to reduce the body mercury burden by an
amount $C_{high}-C_{unstable}$. It is important to observe that a
transition from the low levels of mercury equilibrium point to the
high levels of mercury equilibrium point can be induced by
increasing the body mercury burden by an amount much smaller than
levels corresponding to the difference between high levels and low
levels of mercury burden but to reverse the process one would have
lower body's mercury burden by a large amount.

\begin{figure}[httb]
  \begin{center}
    \subfigure[Mercury body burden. ]
    {\includegraphics[width=0.42\textwidth,height= 0.4\textwidth]{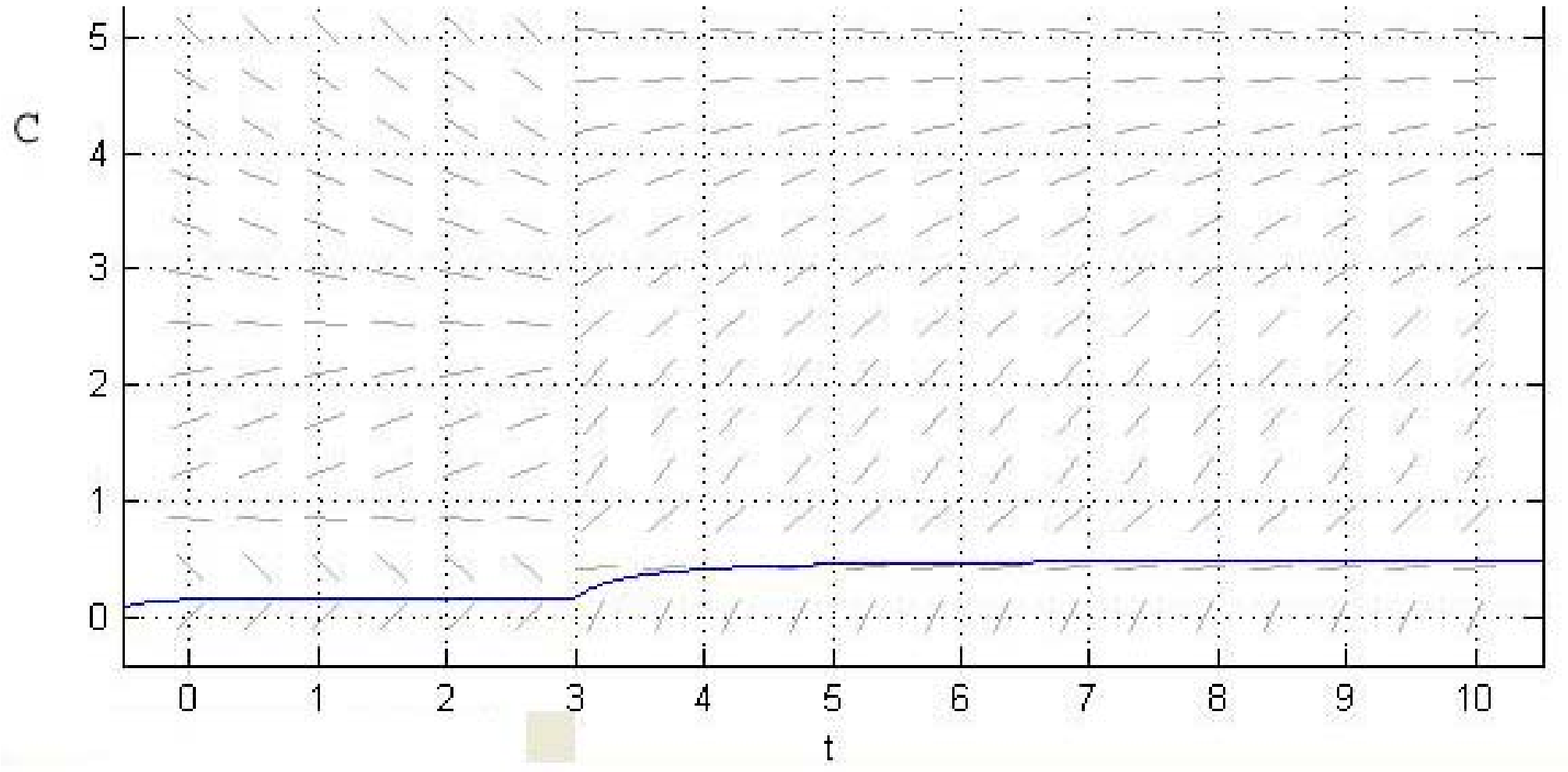}\label{fig1q}}
    \goodgap
    \subfigure[Excretion rate. ]
    {\includegraphics[width=0.42\textwidth,height= 0.4\textwidth]{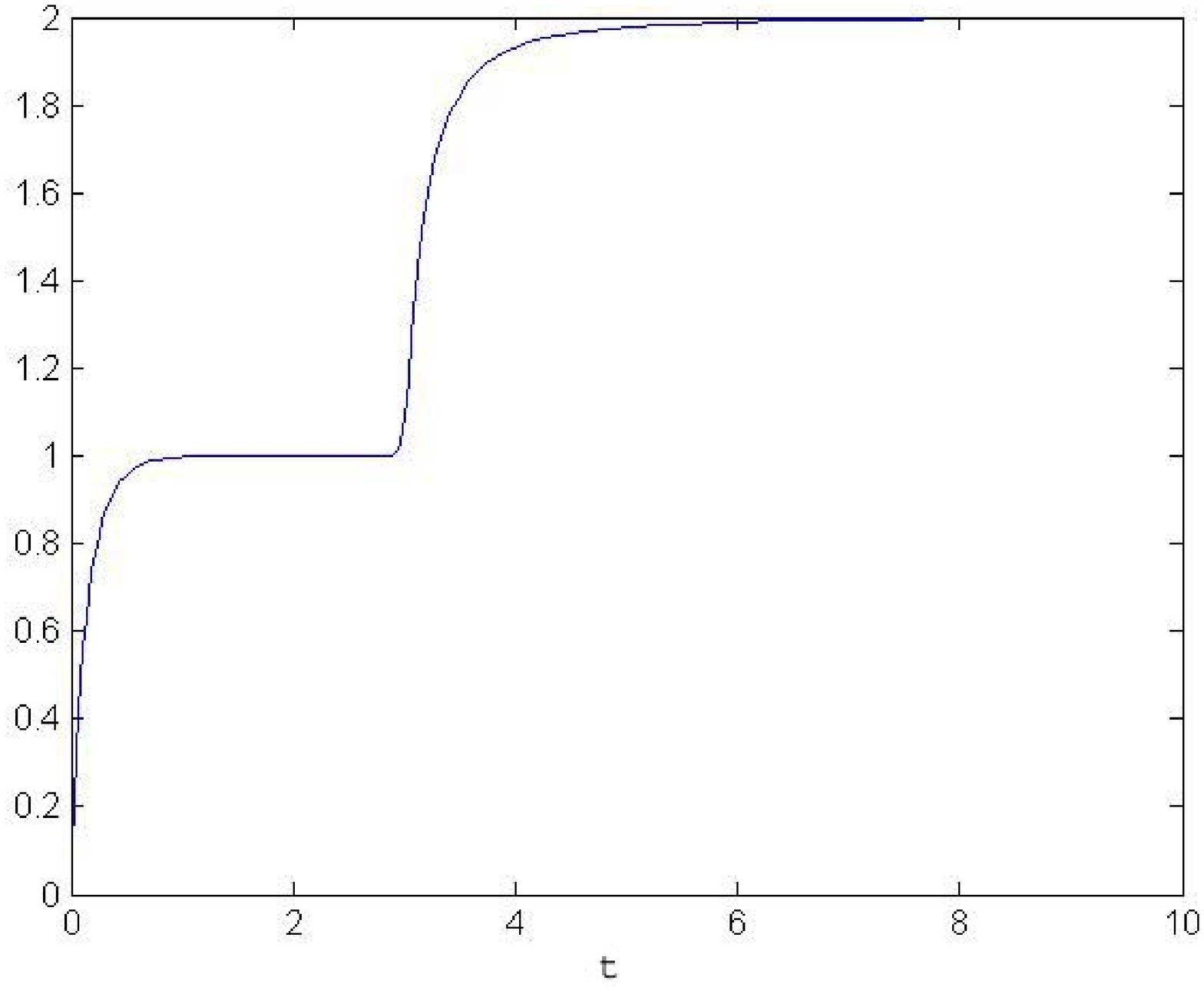}\label{fig2q}}\\
    \caption{At time $3$ the rate of exposure is increased by $\Delta \mu<\Delta \mu_c$.}
    \label{unut100qtip7}
\end{center}
\end{figure}

\begin{figure}[httb]
  \begin{center}
    \subfigure[Mercury body burden. ]
    {\includegraphics[width=0.4\textwidth,height= 0.4\textwidth]{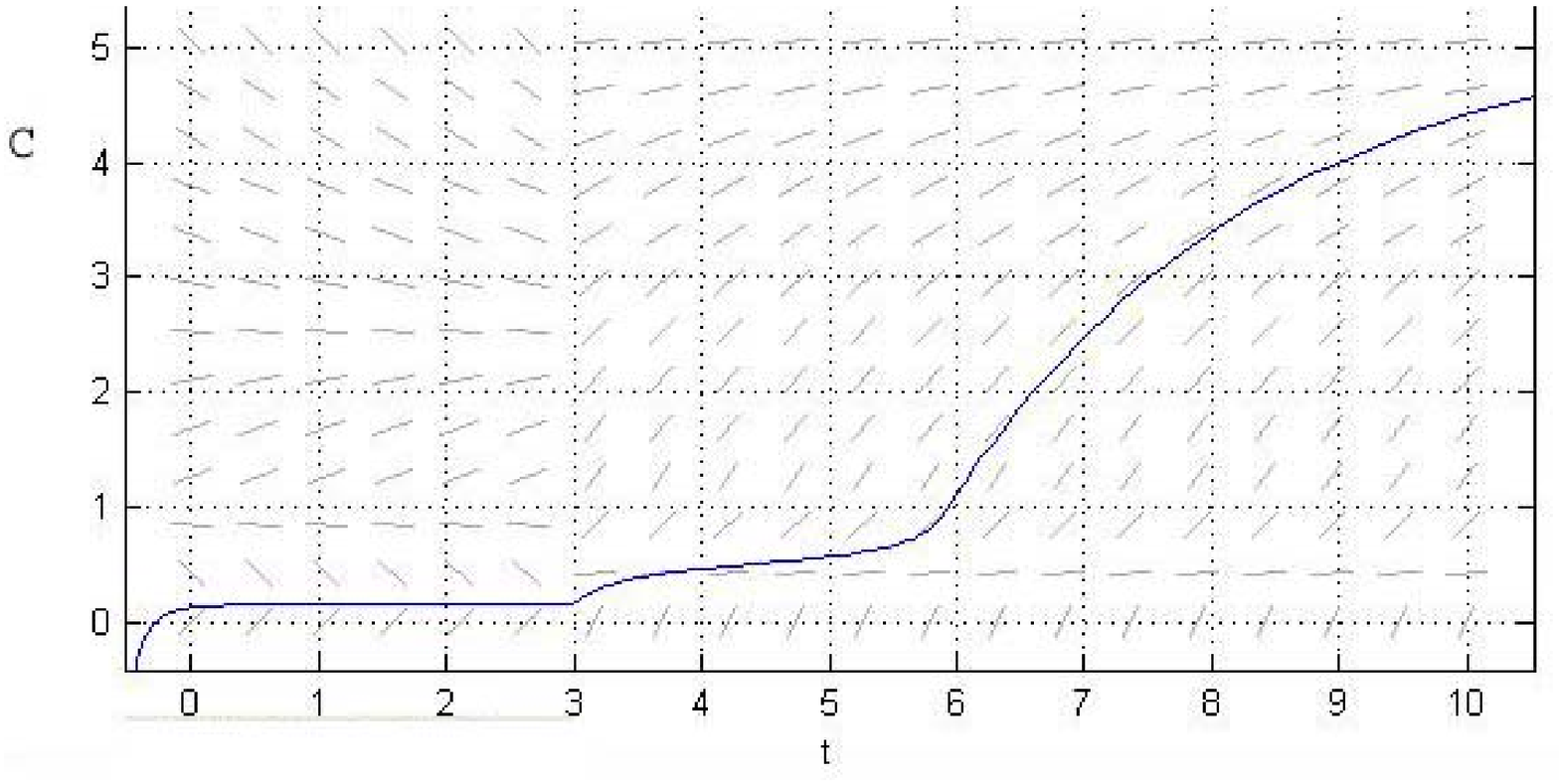}\label{figd1q}}
    \goodgap
    \subfigure[Excretion rate. ]
    {\includegraphics[width=0.4\textwidth,height= 0.38\textwidth]{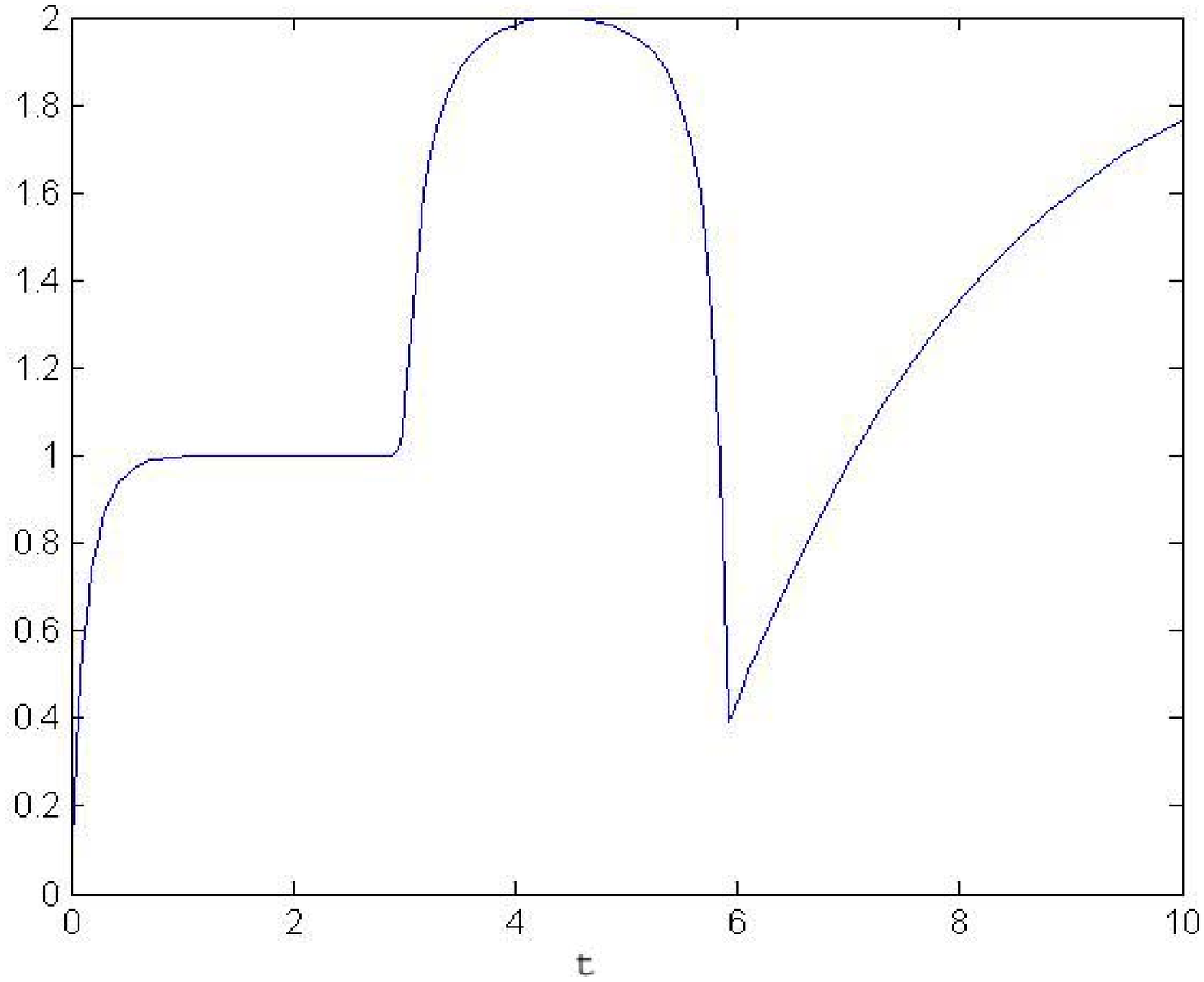}\label{fidg2q}}\\
    \caption{At time $3$ the rate of exposure is increased by  $\Delta \mu>\Delta \mu_c$.}
    \label{unut10d0qstip7}
\end{center}
\end{figure}

\section{Impact of an increase of the exposure rate $\mu$.}
In this section the assumptions behind the model are the following:
at time zero the body burden is assumed to be zero, and the rate of
exposure is assumed fixed. Then time three
 the rate of exposure $\mu$ is increased by a given amount $\Delta \mu$. We observe that if $\Delta \mu$ is
smaller than a critical value$\Delta \mu_c$ (figure
\ref{unut100qtip7}) then the effect effect is a small increase of
the total mercury body burden (figure \ref{fig1q}) and a
corresponding increase of the rate of excretion of mercury (figure
\ref{fig2q}). The mercury body burden remains on the equilibrium
point associated to low levels of mercury and normal abilities to
detoxify.

 If $\Delta \mu$ is larger than
the critical value$\Delta \mu_c$ (figure \ref{unut10d0qstip7} then
the effect is a large accumulation of mercury in the body (figure
\ref{figd1q}). The rate of excretion (figure \ref{fidg2q}) starts
increasing due the increase of mercury exposure, it reaches a
maximum value then starts decreasing due to inhibited abilities to
detoxify. Mercury accumulates in the organs until the equilibrium
points corresponding to a high mercury burden is reached then the
excretion rate increases again to match the exposure rate. In this
second scenario the body burden bifurcates from  low levels of
mercury and normal abilities to detoxify to  high levels of mercury
and inhibited abilities to detoxify.

This theoretical observation based on our toy model is to be put in
relation with s recent study \cite{Jama06} showing that after
placing amalgams on children the urinary excretion of mercury starts
increasing, reaches a peak at year 2, then drop over 40\%
 in the next 5 years to the point where the error bars on the amalgam bearers vs controls
 overlap. This drop in mercury excretion has been interpreted by the
 authors of
 \cite{Jama06} as proof of safety of dental amalgams.
  An other interpretation put forward in \cite{Boy06b} is that children when exposed to mercury vapor for extended
periods of time, slowly lose the ability to excrete mercury in their
urine. Figure \ref{unut10d0qstip7} show that an increase of the rate
of exposure switch off the ability to detoxify and lead to a drop in
the excretion rate after an initial period of increase.

\section{Mercury, a toxic time bomb \cite{BaMe03}.}
Mercury may be involved in several neurological disorders, where
high mercury exposure at one moment or continuously could result in
a reduced ability to detoxify (reduced $k(C)$).

First observe that toxic and neurotoxic effects of heavy metals
\cite{LaJa03} and in particular mercury \cite{ChaL77} are well
known. We refer to \cite{Cla02} for a review of human exposure to
mercury and to \cite{TaMoMa62} for an instance of acute exposure
(Minamata  disease).

Growth delay, reduced locomotion, exaggerated response to novelty,
and densely packed, hyperchromic hippocampal neurons with altered
glutamate receptors and transporters can be induced in mice
following thimerosal challenges that mimic routine childhood
immunizations \cite{HoChLi04}. Thimerosal (ethylmercury) has been
used as a preservative in vaccines since the 1930s \cite{PiCeLo02}.
It has been shown in infant monkeys that \cite{BuShLi05} Thimerosal
crosses the blood-brain barrier and transforms into inorganic
mercury in the central nervous system. For strong evidences of the
link between autism and mercury we refer to \cite{Wind06},
\cite{Palm06} and \cite{Nata06}.

Reduced levels of mercury have been found in first baby haircuts of
autistic children indicating their inability to excrete mercury
\cite{AmBlHa03}. It has been observed in mice \cite{HoChLi04} that
the sensitivity to thimerosal is strain dependent, this may indicate
that  detoxification abilities  in the absence of mercury are
already low in those strains or that the barrier
$C_{unstable}-C_{low}$ is small enough in those strains to be
crossed through injections of low amounts of ethylmercury.

We refer to \cite{SiMoK99} and \cite{Bad52} for the psychometric
evidence that  dental amalgam mercury may be an etiological factor
in schizophrenia. In cerebrospinal fluid of drug-free schizophrenic
patients, a significant decrease in the level of total glutathione
has been observed as compared to controls \cite{DoTrKi00}.

We refer to \cite{NgDe89} and \cite{FiVaKe96} for the association
between body burden mercury level and Parkinson's disease.

There is a significantly greater proportion of males than females
affected by schizophrenia \cite{LeBuMe84}, Parkinson's disease
\cite{WoCuBoLePa04} or autism \cite{GeGe05}, \cite{GeGe06}. This has
to be compared to the fact that at low levels of mercury,
testosterone is known to enhance the neurotoxicity of mercury
whereas oestrogen has protective properties \cite{ClNoSa85}.

It has been observed that blood mercury levels were more than
two-fold higher in Alzheimer's disease patients as compared to
control groups \cite{HoDrGo98} (we also refer to \cite{Ely01}). It
is also known \cite{LeCNa01} that seven of the characteristic
markers of Alzheimer's disease can be produced in normal brain
tissues, or cultures of neurons, by the addition of extremely low
levels of mercury.

It is interesting to observe in that context that chelation
therapies have been proposed for Parkinson's disease \cite{BSER92},
Alzheimer's disease \cite{CuFaXu00}  and autism \cite{Dan05}.
Furthermore high levels of mercury have been detected in autistic
children following chelation therapy with DMSA \cite{BlReBe04},
\cite{BrGeK03}. Furthermore urine excretion of mercury after DMPS
challenge has been shown to be proportional to whole body burden
before chelation \cite{Exra88}.

 A support of the detoxification pathway without the
use of DMSA or DMPS has also been observed to lead a very
significant increase of the rate of excretion of mercury (and other
heavy metals) \cite{YasGor06}. A strong link between detoxification
and recovery from autism has been observed \cite{Dan05},
\cite{YasGor06}.

{\bf Acknowledments.} We would like to thank Boyd E. Haley for
comments which have led to significant changes to this paper, for
bringing to our attention the other factors at play in mercury
toxicity besides Glutathione levels and indicating us references
\cite{Jama06} and \cite{Boy06b}. We would also like to thank two
anonymous referees for valuable and detailed comments and
suggestions.

\def\polhk#1{\setbox0=\hbox{#1}{\ooalign{\hidewidth
  \lower1.5ex\hbox{`}\hidewidth\crcr\unhbox0}}}

\end{document}